\documentclass[prb,showpacs,superscriptaddress,titlepage,amsmath,amssymb,twocolumn]{revtex4-1}

\usepackage{graphicx}
\usepackage{epsfig}
\usepackage{subfigure}
\usepackage{hyperref}
\usepackage{epstopdf}
\usepackage{dcolumn}
\usepackage{bm}
\usepackage{color}

\begin{document}

\title{
A Pure Yang-Mills Description of Anharmonic Phonons
}
\author{J.~M.~Booth}
\email{jamie.booth@rmit.edu.au}
\affiliation{ARC Centre of Excellence in Exciton Science, RMIT University, Melbourne, Australia}

\date{\today}

\pagebreak
\begin{abstract}
It is shown that a unitary theory of interacting phonons with orthogonal polarization vectors can be described by a pure SU(2) Yang-Mills theory. The three orthogonal modes: the two transverse and one longitudinal mode for each value of the momentum must be dotted into the SU(2) generators for interactions to occur in a conveniently unitary form. Without this the commutation relations and the antisymmetry of the Field Strength Tensors results in all scattering vertices cancelling. This suggests that the electron states in strongly anharmonic crystals, which are the phonon source terms, may be non-trivial and consist of a many-body structure which is not describable in terms of single quasiparticles.
\end{abstract}

\maketitle
Anharmoncity in crystal systems is a well-known phenomenon with a long history of study.\cite{Cowley1968} In metal oxides such as VO$_{2}$ and the cuprate superconductors there are many examples of crystal structure transformations suggestive of non-trivial lattice dynamics.\cite{Imada1998} In a previous study\cite{Booth2018_1} it was found that by appropriately grouping the possible spin and electron-hole degrees of freedom, that such materials may have their boson-spinor interactions described by an SU(2) Yang-Mills theory. In that work it was postulated that in some special polar crystal configurations, such as those of the aforementioned oxides, that the transverse phonons may couple to the spins of the electrons and holes. In this work a different tack is taken, the existence of a Yang-Mills theory is demonstrated by assuming: i) that the usual two transverse and one acoustic phonons of a three-dimensional crystal exist, ii) they are described by space- and time-varying vector fields: the polarization vectors, and iii) these phonons can interact with each other.

For a unitary Quantum Field Theory in which all phenomena are described by particle propagation or scattering, we expect that interaction terms entering the Lagrangian can be conveniently represented as polynomial expressions of field operators. For example the scattering of two bosons into a third boson would look something like:
\begin{equation}
\hat{A}\hat{B}\rightarrow \hat{C} = \hat{a}_{c}^{\dagger}\hat{a}_{b}\hat{a}_{a}|\Omega\rangle
\end{equation}
which can then be overlapped with a final state to compute an amplitude. However, for vector bosons, such as phonons, the operators create and annihilate polarization vectors ($\epsilon_{\mu}(p)$), and therefore unitary scattering vertices expressed in this manner must describe the product of two vectors which gives a third vector. A simple way to do this with out loss of generality is for the vertex to consist of a cross product of the incoming vectors. Schematically:
\begin{equation}
\epsilon_{a}(p)\times\epsilon_{b}(p)= \epsilon_{c}(p+q)
\end{equation}
Or in the more suggestive form:
\begin{equation}
\epsilon_{c,\mu}(p+q)=f^{\mu\nu\sigma}\epsilon_{a,\nu}(p)\epsilon_{b,\sigma}(q)
\label{cross_product}
\end{equation}
where here $f^{\mu\nu\sigma}$ is the antisymmetric Levi-Civita symbol (in this work $f$ is used as a label to distinguish it from the polarization vectors), and we expand each polarization vector in a basis of unit vectors describing the $3+1$-dimensional spacetime they inhabit: $\epsilon_{a}(p) \rightarrow \epsilon^{\lambda}_{a,\mu}(p)$, where the $\lambda$ are the basis vectors. Aside: we are addressing the case of phonons here which do not transform as Lorentz vectors as the crystal structure has an underlying orientation, and thus their polarization vectors have no time component. However, the time derivatives are obviously still present, so it is convenient to use the space-time notation with the implication that since $\epsilon^{0}(p) = 0$ then terms of the form $\partial_{\mu}\hat{W}_{0}$ are all zero. Therefore, there is no contribution to the phonon field from a scalar potential, it is a vector potential only. This also means we do not require the Lorentz gauge condition to constrain the polarization vectors, and the Ward-Takahashi identity\cite{Peskin2016} is not required to eliminate the longitudinal mode.


To see how the antisymmetry occurs: for any Vector Field $W_{\mu}$(x) which can be decomposed into normal modes, we can write its dynamics in the form of the Action:
\begin{equation}
S = \int d^{4}x\bigg(-\frac{1}{4}F_{\mu\nu}F^{\mu\nu}\bigg)
\end{equation}
where:
\begin{equation}
F_{\mu\nu} = \partial_{\mu}W_{\nu}-\partial_{\nu}W_{\mu}
\end{equation}
is the curvature form which gives the wave behaviour of the vector modes (i.e. the Field Strength Tensor), and 
\begin{equation}
W_{\mu}(x) = \int \frac{d^{3}p}{{2\pi}^\frac{3}{2}2E_{\mathbf{p}}^{\frac{1}{2}}}\sum_{\lambda}\big[\hat{a}\epsilon^{\lambda}_{\mu}(p)e^{ip_{\mu}x^{\mu}} + \hat{a}^{\dagger}\epsilon^{*\lambda}_{\mu}(p)e^{-ip_{\mu}x^{\mu}}\big]
\label{W-field}
\end{equation} 

That is, the $\partial_{\mu}W_{\nu}$ terms describe the change in the polarization vector's $\nu$ component when incremented in the $\mu$ direction. The time increment obviously gives information on the energy, i.e. the larger the derivative the higher the energy, while the spatial derivatives give information on the wavelength, a smaller spatial derivative obviously will result in a longer wavelength for a normal mode. 

It is an experimental fact that phonon-phonon scattering occurs,\cite{Cowley1968,Ashcroft_2011_1} which renders the potential anharmonic, and thus to describe the field strength of a particular mode at a spacetime point $x$ we must include a contribution which comes from the scattering. That is, the derivative term in the Field Strength Tensor gives the contribution to the polarization vector of the mode from the different momentum states, which is a complex number having norm in the range [0,1], derived from the mode 4-momentum, and we need an extra term to account for interactions. Thus, $F_{\mu\nu}$ for bosons of type $a$ will contain a kinetic factor from the derivative, and also two bosons of type $b$ and $c$ could interact to produce an $a$ boson (for the present, $a,b,c$ are not necessarily different, just a label to capture all of the different possibilities). If we consider a three-dimensional system, we have the usual two transverse and one longitudinal acoustic modes, then for example the contribution to $F_{\mu\nu}$ of the polarization vector of a longitudinal mode inside a particular unit cell may contain terms arising from the scattering of the two transverse phonons. At each spacetime point $x = (t,\mathbf{R})$, where $\mathbf{R}$ labels unit cells, we sum  over the momentum states of each branch, and thus there is no scattering between modes of the same branch (this would be double-counting). 

Therefore, as the contribution to the polarization vector may contain contributions from scattering of the other two modes we must modify the derivative term in the Field Strength Tensor to include this. That is:
\begin{equation}
\partial_{\mu}\hat{W}^{a}_{\nu}\rightarrow \partial_{\mu}\hat{W}^{a}_{\nu}+ig\hat{W}^{a}_{\mu\nu}
\end{equation}
where the $\hat{W}^{a}_{\mu\nu}$ term arises from scattering of the other two modes and the prefactor $g$ describes the amplitude for this scattering to occur (this may be a function of the momentum states, for simplicity this aspect is ingored as it has no bearing on the discussion) and the $i$ just matches the factor pulled down by the derivative. Of course this contribution can be positive or negative, but the field operators contain both creation and annihilation operators. Following our insistence on unitarity, we expect scattering vertices to be polynomial in the field operators, and thus we make the ansatz that we can replace this term by a product of two modes:
\begin{equation}
ig\hat{W}^{a}_{\mu\nu}\rightarrow g\hat{W}^{b}_{\mu}\hat{W}^{c}_{\nu}
\end{equation}
and $b$ and $c$ are not necessarily different to $a$. Putting this into the equation for the Field Strength Tensor:
\begin{multline}
F_{\mu\nu} = \partial_{\mu}W_{\nu}-\partial_{\nu}W_{\mu}\rightarrow\\ (\partial_{\mu}\hat{W}^{a}_{\nu}+g\hat{W}^{b}_{\mu}\hat{W}^{c}_{\nu})- (\partial_{\nu}\hat{W}^{a}_{\mu}-g\hat{W}^{b}_{\nu}\hat{W}^{c}_{\mu})\\
=\partial_{\mu}\hat{W}^{a}_{\nu}-\partial_{\nu}\hat{W}^{a}_{\mu}+g(\hat{W}^{b}_{\mu}\hat{W}^{c}_{\nu}-\hat{W}^{c}_{\mu}\hat{W}^{b}_{\nu})
\label{Field_Strength}
\end{multline}
by virtue of the commutation relations of the boson creation and annihilation operators, and the fact that the $\mu$ and $\nu$ are both summed over, as are the phonon branch labels $a,b,c$. The last term on the right-hand side is just a commutator of the $b$ and $c$ fields, and can be re-arranged to:
\begin{equation}
if^{abc}\hat{W}^{a}_{\mu\nu}=\big[\hat{W}^{b}_{\mu},\hat{W}^{c}_{\nu}\big]
\label{Yang-Mills}
\end{equation}
where again $f^{abc}$ is the anti-symmetric Levi-Civita symbol, i.e. now $a$ \textit{is} different to $b$ which is different to $c$. This is not particularly mysterious, the expression for the Field Strength Tensor of equation (\ref{Field_Strength}) generates all combinations of terms which as a result of the antisymmetry of $F_{\mu\nu}$ gives all possible cross products, equation (\ref{Yang-Mills}) just groups them into a convenient and suggestive form. However, the Vector Fields themselves are bosons, and thus their operators commute, which means that this term is zero if we are taking simple products of the fields. Therefore, if we write a unitary description of the scattering of three dimensional vector fields using the cross-product, we end up with all of the scattering terms cancelling.

To generate a non-zero interaction term we can dot the fields into objects which obey such commutation relations:
\begin{equation}
\hat{W}^{a}_{\mu}(x)\rightarrow \hat{W}^{a}_{\mu}(x)\hat{T}_{a}
\end{equation}
and
\begin{equation}
if^{abc}\hat{T}^{a} = \big[\hat{T}^{b},\hat{T}^{c}\big]
\end{equation}

This is the familiar commutation relation of the generators of a Lie Algebra. As there are three types of acoustic phonon, the two transverse and one longitudinal mode, this commutator defines the algebra of the Lie Group SU(2) and the generators ($\hat{T}^{a}$) are the Pauli matrices. This reveals that the contribution to the Field Strength given by antisymmetric combinations of the derivatives $\partial_{\mu}\hat{W}_{\nu}$ which comes from scattering is the Lie Derivative. Intuitively this makes sense as the curvature form is telling us how rapidly the phonon polarization vectors are changing as a function of spacetime, and the contribution to the polarization vector from scattering (i.e. the \textit{change} in the polarization vector as a result of scattering) is also in the form of a derivative, and carries the same spacetime indices. It is interesting that such a structure arises without any reference to gauge invariance.

Squaring this extended Field Strength Tensor to form an action gives us interaction vertices of the form:
\begin{equation}
F_{\mu\nu}F^{\mu\nu} = g\partial_{\mu}\hat{W}^{a}_{\nu}\hat{W}^{b,\mu}\hat{W}^{c,\nu} - g\partial_{\nu}\hat{W}^{a}_{\mu}\hat{W}^{b,\mu}\hat{W}^{c,\nu} + \dots
\end{equation}
which are polynomials in the field operators, and thus satisfy our requirement for unitarity, and the antisymmetry gives cross products which describe two orthogonal incoming vectors interacting to form a vector orthogonal to the first two, and we have shifted: $\hat{W}^{a}_{\mu} \rightarrow \hat{W}^{a}_{\mu}\hat{T}^{a}$.

Thus we find that representing the three types of phonon field strengths by curvature forms and requiring that they interact generates an SU(2) Yang-Mills theory.\cite{Yang1954} This inclusion of SU(2) generators for interactions between phonons has significant implications for the source terms, i.e. the spinors, which are related to the boson operators by $\partial_{\nu}F^{\mu\nu}=J^{\mu}$. To create vector fields in the form of two-component matrices the spinors to which such terms are coupled must be stacked appropriately. The structures of the source terms and thus the electron-phonon interactions are explored in a different study,\cite{Booth2018_1} where it is found that the electron liquid takes the form of interacting double-stacked 4-component spinors, in which the individual spinors are Weyl spinors consisting of an electron and a hole, i.e. a Nambu spinor. It therefore seems that anharmonicity may not be a consequence of the interactions of single quasiparticles with the lattice, and thus systems in which anharmonicity is present have a richer many-body structure to their spinor states.

The author acknowledges the support of the ARC Centre of Excellence in Exciton Science (CE170100026) and useful conversations with S. Russo, S. Todd, S. Bilson-Thomson and J. Smith. Correspondence and requests for materials should be addressed to JMB, email: jamie.booth@rmit.edu.au
\bibliography{C:/Local_Disk/GWApproximation/Bibliography/library}

\begin{thebibliography}{6}%
\makeatletter
\providecommand \@ifxundefined [1]{%
 \@ifx{#1\undefined}
}%
\providecommand \@ifnum [1]{%
 \ifnum #1\expandafter \@firstoftwo
 \else \expandafter \@secondoftwo
 \fi
}%
\providecommand \@ifx [1]{%
 \ifx #1\expandafter \@firstoftwo
 \else \expandafter \@secondoftwo
 \fi
}%
\providecommand \natexlab [1]{#1}%
\providecommand \enquote  [1]{``#1''}%
\providecommand \bibnamefont  [1]{#1}%
\providecommand \bibfnamefont [1]{#1}%
\providecommand \citenamefont [1]{#1}%
\providecommand \href@noop [0]{\@secondoftwo}%
\providecommand \href [0]{\begingroup \@sanitize@url \@href}%
\providecommand \@href[1]{\@@startlink{#1}\@@href}%
\providecommand \@@href[1]{\endgroup#1\@@endlink}%
\providecommand \@sanitize@url [0]{\catcode `\\12\catcode `\$12\catcode
  `\&12\catcode `\#12\catcode `\^12\catcode `\_12\catcode `\%12\relax}%
\providecommand \@@startlink[1]{}%
\providecommand \@@endlink[0]{}%
\providecommand \url  [0]{\begingroup\@sanitize@url \@url }%
\providecommand \@url [1]{\endgroup\@href {#1}{\urlprefix }}%
\providecommand \urlprefix  [0]{URL }%
\providecommand \Eprint [0]{\href }%
\providecommand \doibase [0]{http://dx.doi.org/}%
\providecommand \selectlanguage [0]{\@gobble}%
\providecommand \bibinfo  [0]{\@secondoftwo}%
\providecommand \bibfield  [0]{\@secondoftwo}%
\providecommand \translation [1]{[#1]}%
\providecommand \BibitemOpen [0]{}%
\providecommand \bibitemStop [0]{}%
\providecommand \bibitemNoStop [0]{.\EOS\space}%
\providecommand \EOS [0]{\spacefactor3000\relax}%
\providecommand \BibitemShut  [1]{\csname bibitem#1\endcsname}%
\let\auto@bib@innerbib\@empty
\bibitem [{\citenamefont {Cowley}(1968)}]{Cowley1968}%
  \BibitemOpen
  \bibfield  {author} {\bibinfo {author} {\bibfnamefont {R.~A.}\ \bibnamefont
  {Cowley}},\ }\href@noop {} {\bibfield  {journal} {\bibinfo  {journal} {Rep.
  Prog. Phys.}\ }\textbf {\bibinfo {volume} {31}},\ \bibinfo {pages} {123}
  (\bibinfo {year} {1968})}\BibitemShut {NoStop}%
\bibitem [{\citenamefont {Imada}\ \emph {et~al.}(1998)\citenamefont {Imada},
  \citenamefont {Fujimori},\ and\ \citenamefont {Tokura}}]{Imada1998}%
  \BibitemOpen
  \bibfield  {author} {\bibinfo {author} {\bibfnamefont {M.}~\bibnamefont
  {Imada}}, \bibinfo {author} {\bibfnamefont {A.}~\bibnamefont {Fujimori}}, \
  and\ \bibinfo {author} {\bibfnamefont {Y.}~\bibnamefont {Tokura}},\ }\href
  {\doibase 10.1103/RevModPhys.70.1039} {\bibfield  {journal} {\bibinfo
  {journal} {Rev. Mod. Phys.}\ }\textbf {\bibinfo {volume} {70}},\ \bibinfo
  {pages} {1039} (\bibinfo {year} {1998})}\BibitemShut {NoStop}%
\bibitem [{\citenamefont {Booth}\ and\ \citenamefont
  {Russo}(2018)}]{Booth2018_1}%
  \BibitemOpen
  \bibfield  {author} {\bibinfo {author} {\bibfnamefont {J.~M.}\ \bibnamefont
  {Booth}}\ and\ \bibinfo {author} {\bibfnamefont {S.~P.}\ \bibnamefont
  {Russo}},\ }\href@noop {} {\  (\bibinfo {year} {2018})},\ \Eprint
  {http://arxiv.org/abs/1808.05769v2} {arXiv:1808.05769v2} \BibitemShut
  {NoStop}%
\bibitem [{\citenamefont {Peskin}\ and\ \citenamefont
  {Schroeder}(2016)}]{Peskin2016}%
  \BibitemOpen
  \bibfield  {author} {\bibinfo {author} {\bibfnamefont {M.~E.}\ \bibnamefont
  {Peskin}}\ and\ \bibinfo {author} {\bibfnamefont {D.~V.}\ \bibnamefont
  {Schroeder}},\ }\href@noop {} {\emph {\bibinfo {title} {{An Introduction to
  Quantum Field Theory}}}}\ (\bibinfo  {publisher} {Westview Press},\ \bibinfo
  {year} {2016})\ pp.\ \bibinfo {pages} {238--244}\BibitemShut {NoStop}%
\bibitem [{\citenamefont {Ashcroft}\ and\ \citenamefont
  {Mermin}(2011)}]{Ashcroft_2011_1}%
  \BibitemOpen
  \bibfield  {author} {\bibinfo {author} {\bibfnamefont {N.~W.}\ \bibnamefont
  {Ashcroft}}\ and\ \bibinfo {author} {\bibfnamefont {D.~N.}\ \bibnamefont
  {Mermin}},\ }\href@noop {} {\emph {\bibinfo {title} {{Solid State
  Physics}}}}\ (\bibinfo  {publisher} {Cengage Learning},\ \bibinfo {year}
  {2011})\ p.\ \bibinfo {pages} {488}\BibitemShut {NoStop}%
\bibitem [{\citenamefont {Yang}\ and\ \citenamefont {Mills}(1954)}]{Yang1954}%
  \BibitemOpen
  \bibfield  {author} {\bibinfo {author} {\bibfnamefont {C.~N.}\ \bibnamefont
  {Yang}}\ and\ \bibinfo {author} {\bibfnamefont {R.~L.}\ \bibnamefont
  {Mills}},\ }\href@noop {} {\bibfield  {journal} {\bibinfo  {journal}
  {Physical Review}\ }\textbf {\bibinfo {volume} {96}},\ \bibinfo {pages} {191}
  (\bibinfo {year} {1954})}\BibitemShut {NoStop}%
\end{thebibliography}%
\end{document}